\documentclass[aps,prb,twocolumn,superscriptaddress,floatfix,longbibliography]{revtex4}
\usepackage{amsfonts}
\usepackage{amssymb}
\usepackage{graphicx}
\usepackage{dcolumn}
\usepackage{bm}
\usepackage{amsmath}
\usepackage[colorlinks,linkcolor=magenta,citecolor=blue,urlcolor=blue]{hyperref}
\usepackage{changes}

\begin{document}

\title{A general approach to the exact localized transition points of 1D mosaic disorder models}

\author{Yanxia Liu}
\email{yxliu-china@ynu.edu.cn}
\affiliation{School of Physics and Astronomy, Yunnan University, Kunming 650091, PR China}
\date{\today }

\begin{abstract}

In this paper, we present a general  correspondence between the mosaic and non-mosaic models,
which can be used to obtain the exact solution for the mosaic ones. This relation holds
not only for the quasicrystal models, but also for the Anderson models.
Despite the different localization properties of the specific models, this relationship shares a unified form.
Applying our method to the mosaic Anderson models, we find that there is a discrete set of  extended states.
At last, we also give the general analytical mobility edge for the mosaic slowly varying potential models and
the mosaic Ganeshan-Pixley-Das Sarma models.

\end{abstract}

\maketitle


\section{Introduction}

The Anderson localization (AL) plays a fundamental role in condensed matter physics \cite{anderson1958absence},
which is conventionally discussed in disordered systems. There are roughly two types of disordered potentials:
random disorders and quasi-periodic potentials. The random disorders are also referred as Anderson models,
where all states are exponentially localized in one and two dimensions with infinitesimal disorder strength,
and the localized and extended states are separated by energy-dependent mobility edge (ME) in three dimensions
\cite{abrahams1979scaling,lee1985disordered,evers2008anderson,Thouless72}.
The one-dimensional(1D) quasi-periodic models have various forms, such as short-range (long-range) hopping processes \cite{biddle2011localization,biddle2010predicted,ganeshan2015nearest,li2016quantum,li2017mobility,li2018mobility,DengX}
, modified quasiperiodic potentials\cite{sarma1988mobility,sarma1990localization,YCWang2020}, and
\sout{some} various extensions of AA models \cite{Kohmoto1983,Ceccatto,Zhou2013,Cai,DeGottardi,Kohmoto2008,WangYC-review,Chandran,Chong2015}. These models support
localization transitions or MEs, and some of them can be obtained exactly.
Localization transition in quasiperiodic systems has attracted increasing interest both theoretically and experimentally in
recent years \cite{luschen2018,Aubry1980,Kohmoto1983,Thouless1988,roati2008,An2018,An2021}.
Some of these models have been realized in experimental platforms such as ultracold gases.

Recently, some 1D mosaic quasicrystals with exact MEs are devised \cite{YCWang2020,Dwiputra2022,Gong2021,Liu2021},
whose MEs cannot be exactly solved via traditional methods.
Luckily, the Avila's global theory \cite{Avila2015,Avila2017} provides us with a 
mathematically rigorous tool to solve this kind of models exactly.
 However, there are two setbacks in the application of Avila’s global theory: one is the models must
 satisfy extra restrictions while the other one is Avila’s global theory is very hard for
many physicists without the specific knowledge in this mathematical area. Meanwhile, Avila’s global theory
 is powerless in dealing with random disorder systems.

There are already many results of mosaic disorder systems have been obtained\cite{YCWang2020,Dwiputra2022,Gong2021,Liu2021}.
However, our knowledge of the mosaic models is still far from comprehensive.
Whether there exists any relevance between the 1D non-mosaic models and mosaic models, and
whether there exist any new states only in mosaic models, still remains unanswered and will
be explored in the following sections.
In this paper, we discovered a general correspondence between the non-mosaic models and
mosaic models which provides us a simple and elegant tool solving the mosaic models exactly
when the corresponding non-mosaic models can be solved exactly.

This paper is organized as follows. In section II, we first introduce the mosaic models and present the
general correspondence between the non-mosaic models and mosaic models.
In section III, we apply our methodology to some specific cases.
In the subsection A, we verify our method with two type models: the mosaic AAH models and the
mosaic Wannier-Stark models, which have been studied by Avila's global theory. Then, in subsection B, we
analytically study the localization properties of the mosaic Anderson models and
find that there exists a set of discrete eigenenergies for extended states, which is independent of the
disorder strength.  In subsection C, we also take the 1D  mosaic slowly varying potential models as a example and give
the general MEs for arbitrary $\kappa$. In subsection D, we give the analytical MEs of the mosaic
Ganeshan-Pixley-Das Sarma models. Final section is a summary.


\section{correspondence between mosaic and non-mosaic models}

The general 1D mosaic models can be written as

\begin{eqnarray}\label{ham-1}
H &=&t\sum_{j} (c^{\dagger}_{j+1}c_{j}+H.c.)+\sum_j V_j n_{j},
\end{eqnarray}
with petential
\begin{equation}\label{potential}
V_j=
\begin{cases}
\lambda \Delta_m,\ \  j=m\kappa,  \\
0, \ \textrm{otherwise},
\end{cases}
\end{equation}
where $c^{\dagger}_{j}$ ($c_{j}$) and  $n_{j}=c^{\dagger}_{j}c_{j}$ are the creation (annihilation) operator
and the number operator at site $j$, $t$ is the hopping coefficient (for convenience, we set $t=1$ as
the energy unit), and $\lambda \Delta_m$ is the potential, which takes place at every $\kappa$ sites.
The explicit forms $\Delta_m$ can make it disordered or not.
We take the wave function as $\Psi =\sum_{j}\psi_{j}\hat{c}_j^{\dagger}|0\rangle$.
The sch\"ordinger equation $H\Psi=E\Psi$ then takes the form of

\begin{eqnarray} \label{LEGs}
\left\{
\begin{array}{c}
\psi_{\kappa (m-1)}+\psi_{\kappa  (m-1)+2}=E\psi_{\kappa (m-1)+1},  \\
\ldots, \\
\psi_{\kappa m-2}+\psi_{\kappa m}=E\psi_{\kappa m-1} , \\
\end{array} %
\right.
\end{eqnarray}
and
\begin{eqnarray}
\psi_{\kappa m-1}+\psi_{\kappa m+1}+\lambda \Delta_m \psi_{\kappa m}=E\psi_{\kappa m} \label{eigen_kkm},
\end{eqnarray}
Here we demand  $m>1$ and ignore the boundary conditions. Eqs. \eqref{LEGs} can give us
\begin{eqnarray}
 \psi_{\kappa m-1} &=&\frac{1}{a_{\kappa} }  \psi_{\kappa (m-1)} + \frac{a_{\kappa-1}}{a_{\kappa} } \psi_{\kappa m} \ \label{tp1},
\end{eqnarray}
and taking $m\rightarrow m+1$ in Eqs. \eqref{LEGs} give us
\begin{eqnarray}
 \psi_{\kappa m+1} &=&\frac{1}{a_{\kappa} }  \psi_{\kappa (m+1)} + \frac{a_{\kappa-1}}{a_{\kappa} } \psi_{\kappa m}  \label{tp2},
\end{eqnarray}
where
\begin{eqnarray} \label{akappa}
 &&a_{\kappa}  (E)=\\\notag
&& \frac{1}{\sqrt{E^2-4}} ((\frac{E+\sqrt{E^2-4}}{2})^{\kappa}-(\frac{E-\sqrt{E^2-4}}{2})^{\kappa}).
\end{eqnarray}
The detailed derivation of Eq. \eqref{tp1} and Eq. \eqref{tp2} is given in the Appendix. We then substitute
Eq. \eqref{tp1} and Eq. \eqref{tp2} into Eq. \eqref{eigen_kkm}
\begin{eqnarray}
\psi_{\kappa (m-1)} +\psi_{\kappa (m+1)} +\lambda a_{\kappa}  \Delta_m \psi_ {\kappa m}
=(a_{\kappa}E-2a_{\kappa-1})\psi_{\kappa m},\notag \\\label{eigen_pm}
\end{eqnarray}
which is a difference equation only including the wave function at sites $j=\kappa m$.
When $\kappa=1$, Eq. \eqref{eigen_pm} is the Schr\"odinger equation of the non-mosaic model, where
the localization-delocalization transitions can occur for disordered $\Delta_m$. Normally, the regions
of extended states are determined by
\begin{equation}
f(\lambda,E)<1,~~\text{or}~~ f(\lambda,E)=1.\label{ldl}
\end{equation}%
The explicit choice depends on the model considered.
The $f(\lambda,E)$ is a function of $\lambda$ and $E$, whose forms depend on the concrete potential $\Delta_m$.
When $\kappa\neq 1$, the extended states only  appear at
\begin{eqnarray}\label{lddl}
f(a_{\kappa} \lambda ,a_{\kappa}E-2a_{\kappa-1})<1,
\end{eqnarray}
or\\
\begin{eqnarray}
f(a_{\kappa} \lambda ,a_{\kappa}E-2a_{\kappa-1})=1.
\end{eqnarray}%
That means once the localization-delocalization transition points of the models with $\kappa=1$ is known,
the localization-delocalization transition points of the models with $\kappa\neq 1$ can be obtained by simple substitution
$\lambda \rightarrow a_{\kappa} \lambda$; $E \rightarrow a_{\kappa}E-2a_{\kappa-1}$.
This provides a straightforward way solving the mosaic models and show the
mechanism of multiple mobility edges in the mosaic models.

Since $a_{\kappa}(E)$ is a function of $E$ to the $\kappa-1$ degree,
$a_{\kappa}E-2a_{\kappa-1}$, which appears at the "energy" position of  \eqref {eigen_pm},
is a $\kappa$th order polynomial of $E$.
If $a_{\kappa}E-2a_{\kappa-1}$ has $L$ solutions, the number of eigenenergies $E$ for Hamiltonian \eqref {ham-1}
are $N=\kappa L$, where $N$ is the lattice length and  $L$ is the quasi-cell number.
Eq. \eqref{eigen_pm} thus gives a complete set of solutions of Hamiltonian \eqref {ham-1}.
The forms of Eq. \eqref{eigen_pm}
with different $\kappa$ are essentially the same, as the $a_{\kappa}(E)$ is independent of the
explicit forms of potentials. Thus, the eigenstates of the systems with different $\kappa$ have
the same structure. So, the cases $\kappa>1$ will not exhibit new states to $\kappa=1$.
We can also see that if the non-mosaic models have the same transition points,
the corresponding mosaic models will have same transition points, regardless the non-mosaic models
being disorder or not.

\section{Applications}

\subsection{The mosaic AAH models and mosaic Wannier-Stark models}

To show the validity of our method, we first consider the mosaic AAH models with
$\Delta_m= 2 \cos (2\pi\alpha m)$ , which host multiple MEs and has been studied in ref \cite{YCWang2020}.
Here $\alpha=(\sqrt{5}-1)/2$ is an irrational number.
The MEs have been exactly solved by applying the Avila's global theory.
The Avila's global theory is an important mathematical result, but is not friendly to the physics researcher who
is not a mathematician and doesn't know much specific mathematical knowledge. Now we deal this model
with a much simpler method, which does not involve complex mathematics.

When $\kappa=1$, Eq. \eqref{eigen_pm} is the simpliest quasiperiodic model (AAH model) and exhibits self-duality symmetry.
Thus the localization-delocalization transition points of this model can be exactly determined by a
self-duality condition \cite{Aubry1980,Jitomirskaya1999},
which can be expressed as $\lambda=1$ and the extended region is  $\lambda<1$,  i.e. $f(\lambda,E)<1$ with $f(\lambda,E)=|\lambda|$.
Substitutting $\lambda$ with $a_\kappa\lambda$, the localization-delocalization points for the cases $\kappa>1$
read
\begin{eqnarray}
|a_\kappa\lambda|=1\label{mob1},
\end{eqnarray}
which is in agreement with the result from the Avila's global theory \cite{YCWang2020} .


This method can also be applied to the mosaic Wannier-Stark models, which can be
described by Eqs. \eqref{ham-1} and \eqref{potential} with $\Delta_m =m$ (a tilted potential).
For $\kappa=1$, all the eigenstates are extended when $|\lambda|<2$, i.e., $f(\lambda,E)<1$ with
$f(\lambda,E)=\lambda-1$. This model has the same characteristics with AAH model: the transition
points are independent of the energy(no ME). Do the same thing, substitutting $\lambda$ with
$a_\kappa\lambda$, then we get that for the cases $\kappa > 0$, extended states exist when
\begin{eqnarray}
|a_\kappa\lambda|>2\label{mob2},
\end{eqnarray}
which is also verified with Avila's global theory \cite{Dwiputra2022}.

\subsection{The 1D  mosaic Anderson models}

Here we use our method to study 1D mosaic Anderson models, which can be described by
Eqs. \eqref{ham-1} and  \eqref{potential} with $\Delta_m =w_m$,
where $w_m \in \left[-1,1\right] $ is uniformly distributed random variable.
For the case $\kappa=1$, the infinitesimal disorder strength $\lambda$ makes all eigenstates \sout{are} localized,
that is to say, only when  $\lambda=0$, the states are extended, and it is shown in Eq. \eqref{ldl} with
$f(\lambda,E)=\lambda+1=1$, which is independent of $E$. Now, we substitute
$\lambda$ with $a_\kappa\lambda$ for the case
$\kappa \neq1$, where the extended states appear at $a_\kappa\lambda=0$.
In non-trivial case $\lambda\neq 0$,  the extended states are located at
 \begin{eqnarray}
a_\kappa(E)=0.\label{akappa0}
\end{eqnarray}
The solutions of Eq. \eqref{akappa0} are discrete points, which are independent of the $\lambda$.
Before solving the Eq. \eqref{akappa0}, we define $E=2\cos\theta$ in Eq. \eqref{akappa} for simplicity.
After some algebras, we can get $a_\kappa(2\cos\theta)=\sin(\kappa\theta)/\sin\theta. $
The solutions of $a_\kappa(2\cos\theta)=0$ are given by $\theta=\pm l\pi/\kappa$ with
$l=1,2,\ldots,\kappa-1$. There are $2(\kappa-1)$ different solutions.  The solutions $\theta$ and
$-\theta$ correspond to the same $E$, so there are $\kappa-1$ independent roots of
 $a_\kappa(E)=0$, which can be written as
\begin{equation}\label{SL}
E=2\cos(l\pi/\kappa), ~~l=1,2,\ldots,\kappa-1.
\end{equation}
For $\kappa=2$, the extended states appear at energy $E=0$, while for $\kappa=3$ and $\kappa=4$,
the extended states appear at energy $E=\pm1$ and $E=\pm\sqrt{2},~0$, which is independent of $\lambda$.
We can find that if $\kappa\neq 0$, there exist some
extended states with discrete energies, which do not exist in Anderson models or quasiperiodic lattice models.

The quasiperiodic lattices are some intermediates  between periodic and random lattice, which can realize the crossover
of physical property from periodic to random lattices by change the strength of the quasiperiodic potential.
Actually, the mosaic Anderson models can be also seen as some intermediates between periodic and random lattices.
It becomes the standard  Anderson models when $\kappa=1$,
and becomes the periodic model when $\kappa=\infty$
 (when the lattice site $N$ is finite, $\kappa>N$).
The mechanism of the crossover of this model is that the extended states comes into play with increasing $\kappa$,
 which are unaffected by $\lambda$.

\begin{figure}[tbp]
\includegraphics[width=0.45\textwidth]{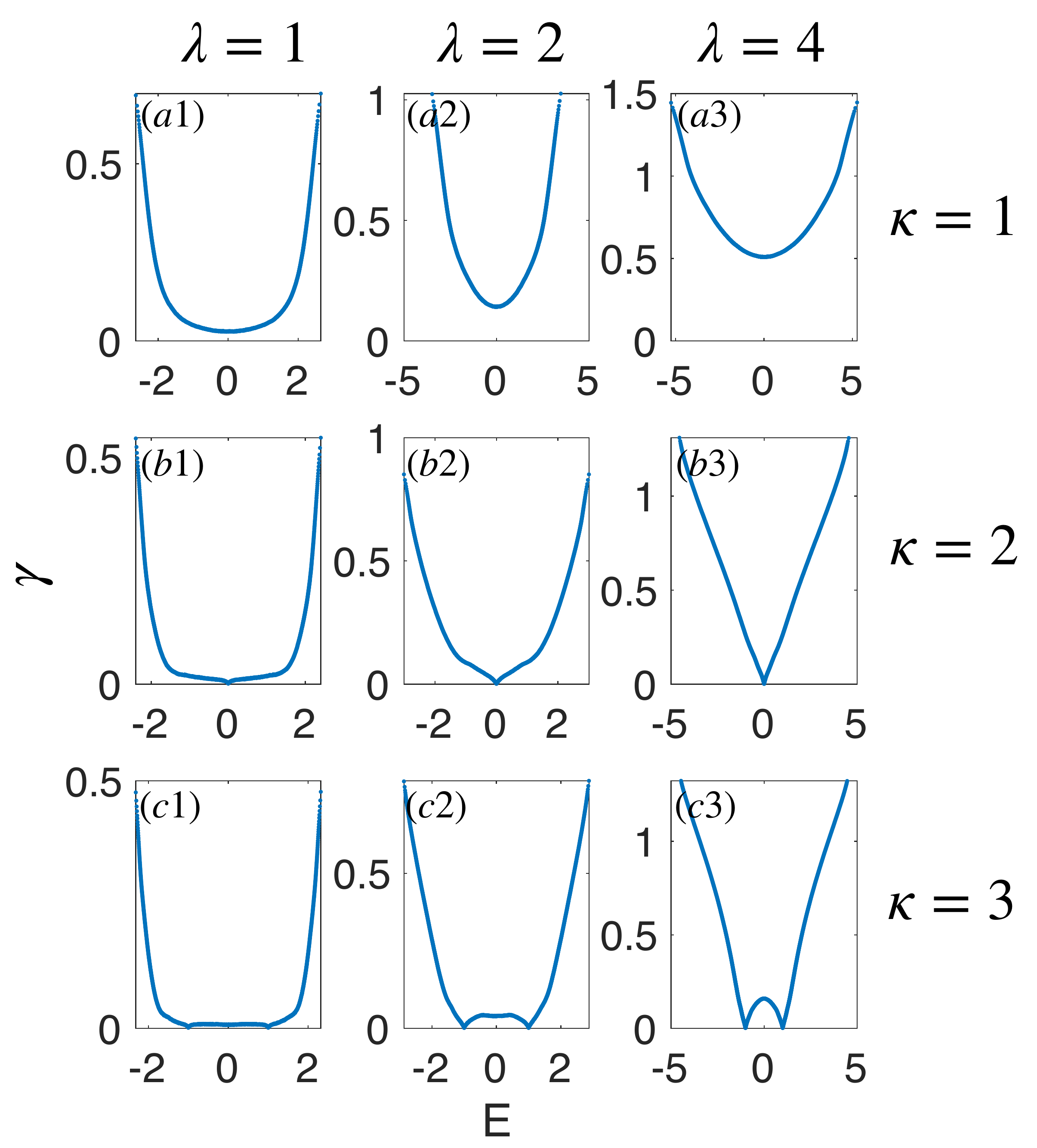}
\caption{ Numerical results for the LE of all eigenenergies with (a1)-(a3) $\kappa=1$, lattice length is $N=2000$,
 (b1)-(b3) $\kappa=2$, $N=2000$,  and (c1)-(c3) $\kappa=3$ $N=2001$, respectively.
 The value of parameter $\lambda$ in each column is the same: (a1)-(c1) $\lambda=1$,
(a2)-(c2) $\lambda=2$, and (a3)-(c3) $\lambda=4$, respectively. All results are the average of $200$ random samples.}
\label{fig1}
\end{figure}

To conceptualize this, we demonstrate numerical results of Lyapunov exponent (LE).
The LE can be numerically calculated via
\begin{equation}
\gamma \left(
E\right) =\ln \left( \max \left( \theta _{i}^{+},\theta _{i}^{-}\right)
\right) , \label{NLE}
\end{equation}
where $\theta _{i}^{\pm }\in \mathbb{R}$ denote eigenvalues of
the matrix
\begin{equation}
\mathbf{\Theta =}\left( T_{N}^{\dag}
T_{N} \right) ^{1/(2N)} . \label{NLE2}
\end{equation}
with the transfer matrix
\begin{equation}
T_{N}\left( E\right) =\prod_{j=1}^{N}T^{j}=\prod_{j=1}^{N}\left(
\begin{array}{cc}
E-V_{j}& -1 \\
1 & 0%
\end{array}%
\right),
\end{equation}%
where $N=\infty$ as required by the definition of  LE.
The LE is a non-negative number ($\gamma \geqslant 0$) which characterize the localization
properties of the eigenstates. A localized state of disorder systems can be expressed as
\[
\left\vert \psi_{j}\right\vert \propto e^{- \gamma \left\vert j-j_{0}\right\vert
},
\]
where $j_{0}$ is the localization center. It is clear that $\psi_{n}$ is a localized state with $\gamma>0$
and an extended state with $\gamma=0$.

In Figs. \ref{fig1}(a1)-(a3), we display the LE of all eigen-energies of the Anderson model ($\kappa=1$)
with $\lambda=1$, $2$, and $4$, respectively. We can see that for $\gamma>0$, all eigenstates
are localized, and for fixed $\lambda$,  the LE increases with the increase of absolute value of
eigenenergies $|E|$ and the minimum LE is at $|E|=0$.  As $\lambda$ increases, the value of minimum LE
increases. Figs. \ref{fig1}(b) and (c) show the LE $\gamma$ for the case $\kappa=2$ and $\kappa=3$  with different $\lambda$.
When $\kappa=2$, the LE $\gamma=0$ with eigenenergies $E=0$ and $\gamma>0$ otherwise,
which is independent of $\lambda$.
When $\kappa=3$, the LE $\gamma=0$ with eigenenergies $E=\pm 1$ and $\gamma>0$ otherwise.
In other word,  only the eigenstates $E=0$ for $\kappa=2$ and
 $E=\pm 1$ with $\kappa=2$ are extended, which is consistent with our analytical result.

\subsection{The 1D  mosaic  slowly varying potential models}
In this part, we extend our method to the mosaic slowly varying potential model,
 which is another famous quasiperiodic models described by
Eqs. \eqref{ham-1} and \eqref{potential} with $\Delta_m =\cos(\pi \alpha m^{\nu})$,
$0<\nu<1$. For the case $\kappa=1$,  the extended states are clustered in $|E|<2-\lambda$
with $\lambda<2$, which can be obtained by asymptotic semiclassical WKB-type theory
\cite{sarma1988mobility,sarma1990localization}. When $\lambda>2$, all states are localized.
Applying Eq. \eqref{ldl}, the extended region is $f(\lambda,E)<1$ with $f(\lambda,E)=|E|-1+\lambda$.
The Eq. \eqref{lddl} provides us the condition for extended states with $\kappa>1$.
Substituting $\lambda$ with $a_\kappa\lambda$ and $E$ with $a_\kappa E-2 a_{\kappa-1} $, we get
\begin{eqnarray}\label{SV1}
&&f(a_{\kappa} \lambda ,a_{\kappa}E-2a_{\kappa-1}) \notag, \\
&=&|a_\kappa E-2 a_{\kappa-1}|-1+a_\kappa\lambda<1,
\end{eqnarray}
with $a_\kappa\lambda<2$. When $\kappa=2$, $a_{2}=E$; $a_{1}=1$, then we get
$f(a_{2} \lambda ,a_{2}E-2a_{1})=|E^2-2|-1+E\lambda<1$ and $E\lambda<2$ are the extended regions.
When $\kappa=3$, $a_{3}=E^2-1$, $f(a_{3} \lambda ,a_{3}E-2a_{2})=|E^3-3E|-1+(E^2-1)\lambda<1$ and
 $(E^2-1)\lambda<2$ are the extended regions.
The results of the case $\kappa=2$ and $3$ are agreement with the result in \cite{Gong2021}
([\cite{Gong2021}] doesn't give a general result for arbitary $\kappa$).

\subsection{The mosaic Ganeshan-Pixley-Das Sarma models}

\begin{figure}[t]
\hspace*{-0.5cm}
\includegraphics[width=0.52\textwidth]{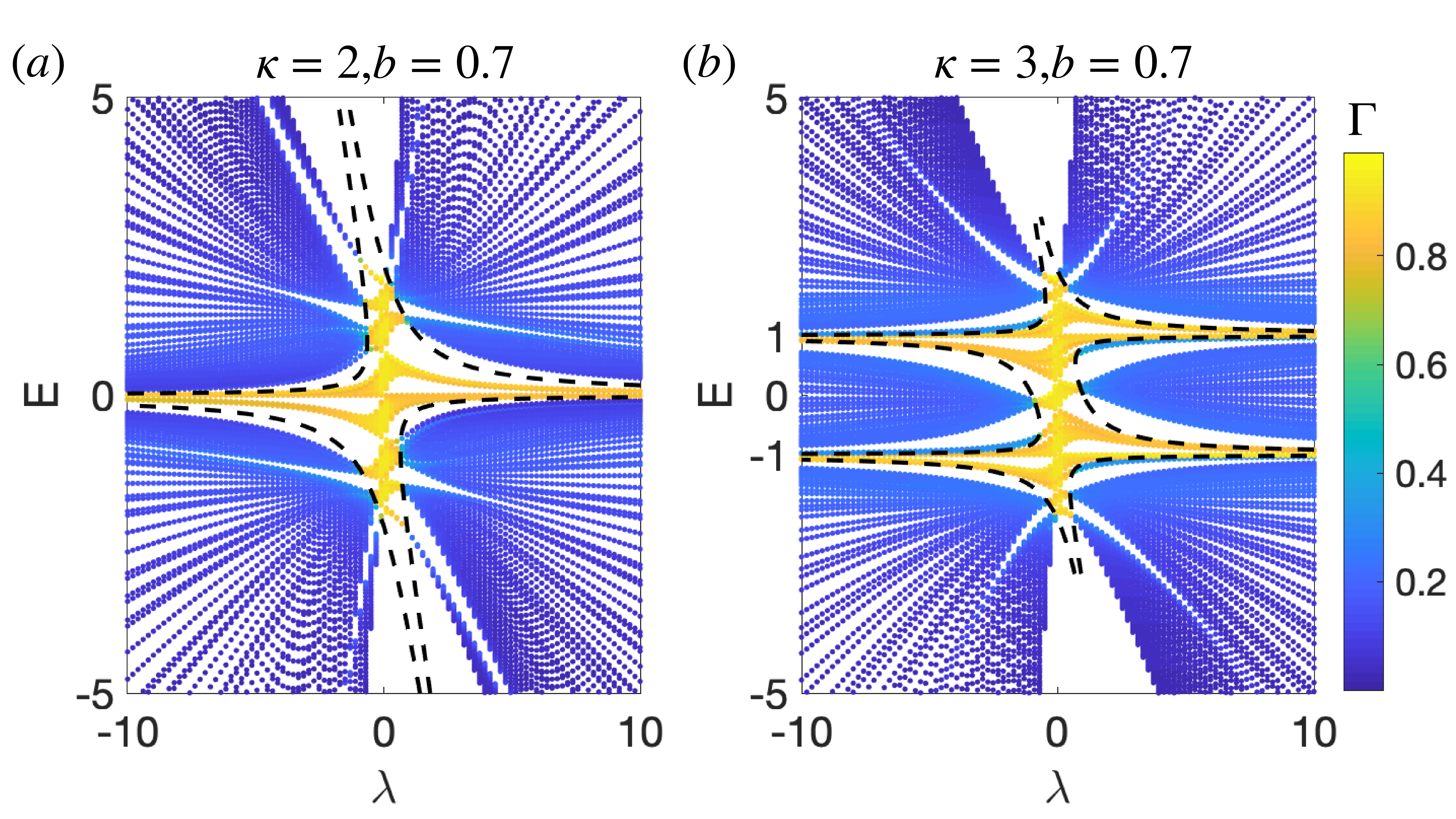}
\caption{\label{fig2}
Fractal dimension $\Gamma$ of different eigenstates as a function of the corresponding eigenvalues and
quasiperiodic potential strength $\lambda$ for (a) $\kappa=2$ with size $N=2\times 233$ and (b) $\kappa=3$
with size $N=3\times 233$. Without loss of generality, we set $\theta=0$. The black
dashed lines represent the MEs given in Eq.~\eqref{ME2} and \eqref{ME3}.}
\end{figure}

Now, we consider the mosaic Ganeshan-Pixley-Das Sarma model
 \begin{equation}
\Delta_m= \frac{ 2\cos (2\pi\alpha m+\theta)}{1-b\cos (2\pi\alpha m+\theta) }, \label{Vfrac}
\end{equation}
where $b\in (-1,1)$ and $\theta$ is the phase offset. The case with $\kappa=1$ is the first quasiperiodic model where the analytic form of
mobility edges are obtained by looking for the self-dual point \cite{ganeshan2015nearest} .
The extended region is
\begin{equation}\label{GPD}
2 \text{sgn}(\lambda)(1-|\lambda|)<bE,
\end{equation}
where $\text{sgn}(~)$ is sign function, then we get
\begin{equation}\label{GPDf}
f(\lambda,E)=2 \text{sgn}(\lambda)(1-|\lambda|)-bE+1<1.
\end{equation}
When $\kappa\neq1$, substituting $\lambda$ with $a_\kappa\lambda$ and $E$ with $a_\kappa E-2 a_{\kappa-1} $,
we get the extended regions:
\begin{eqnarray}\label{GPDk}
&&f(a_{\kappa} \lambda ,a_{\kappa}E-2a_{\kappa-1})  \\
&=&2 \text{sgn}(a_\kappa\lambda)(1-|a_\kappa\lambda|)-b(a_\kappa E-2 a_{\kappa-1})+1<1\notag,
\end{eqnarray}
that is, the MEs are
\begin{eqnarray}\label{GPDme}
2 \text{sgn}(a_\kappa\lambda)(1-|a_\kappa\lambda|)=b(a_\kappa E-2 a_{\kappa-1}),
\end{eqnarray}
It is obvious that Eq. \eqref{GPDme} with $b=0$ is identical to the MEs of the mosaic AA model:
\begin{eqnarray}\label{ME1A}
|\lambda a_{\kappa}|=1.
\end{eqnarray}
And the LE \eqref{GPDme} with $\kappa=2$ and $\kappa=3$ can lead to
\begin{eqnarray}\label{ME2}
\lambda  =\frac{(2-E^2)b\pm2}{2E},
\end{eqnarray}
and
\begin{eqnarray}\label{ME3}
\lambda=\frac{b(3E-E^3)\pm2}{E^2-1} .
\end{eqnarray}
Equations~\eqref{ME2} and \eqref{ME3} shows the energy boundary of localized and extended states.

To characterize the ME, we introduce the fractal dimension of  an eigenstate \cite{li2016quantum,li2018mobility,li2017mobility}.
which is defined as
\begin{equation}
\Gamma=-\lim_{N\rightarrow \infty}\ln ({\rm IPR})/\ln N, \label{FD}
\end{equation}
where the inverse participation ratio ${\rm IPR}= \sum_{j}\left\vert \psi_{j}\right\vert^{4}$.
The fractal dimension $\Gamma$ approaches $0$ for a localized state and approaches $1$
for an extended state. Figure \ref{fig2} shows the distribution of the spectra versus $\lambda$ with $b=0.7$
for $\kappa=2$ and $3$, respectively. The different color of each spectrum represents the fractal dimension $\Gamma$.
The dashed lines in the Fig. \ref{fig2}~(a) and (b) represent the MEs~\eqref{ME2} and \eqref{ME3} for $b=0.7$.
The eigenstates corresponding to the spectra surrounded by dashed lines are extended,
and outside the four ($\kappa=2$) or six ($\kappa=3$) dashed lines are localized.
It is shown that the analytical MEs agrees well with numerical results from fractal dimension $\Gamma$ and spectra.
Moreover, we can also obtain the emergence of multiple mobility edges in these models,
which can be regarded as $f(a_{\kappa} \lambda ,a_{\kappa}E-2a_{\kappa-1})$ for $\kappa>1$ being
a power function of $E$.

For $\lambda\rightarrow \infty$, the extended eigenstates are concentrated at $E=0$
for $\kappa=2$ and $E=\pm 1$ for $\kappa=3$, which is independent of $b$.
Something interesting is, when $\lambda\rightarrow \infty$, the concentration of spectral of extended states
is identical to that of Anderson model, which is irrelevant of $\lambda$.

\section{Summary}

In this paper, we have proposed a new way to deal with the mosaic disorder systems via the
decoupling the schr\"odinger equations. Such method not only establishes a general correspondence between
1D non-mosaic disorder models and mosaic disorder models, but also can give exact solutions of MEs to
a variety of  models, such as the mosaic slowly varying potential models and  the mosaic Ganeshan-Pixley-Das Sarma models.
With this method, we strictly demonstrate that there is a discrete set of extended states in 1D mosaic Anderson models
and revealed the mechanism of the emergence of multiple mobility edges.

\begin{acknowledgments}

We thank Y.-S. Cao for useful discussion.
  \end{acknowledgments}

\appendix

\section{Decoupling difference equation}

The potentials of the mosaic quasi-periodic models take place with fixed site interval $n=\kappa m$.
As one can infer from Eqs. \eqref{LEGs} and \eqref{eigen_kkm}, the wave functions for the lattice sites $n=\kappa m$ can be decoupled from
the wave functions that at other sites $n\neq \kappa m$. The difference equation involving potential is
\begin{eqnarray}
\psi_{\kappa m-1}+\psi_{\kappa m+1}+\lambda \Delta_{ m} \psi_ {\kappa m}=E\psi_{\kappa m} \label{eigen_km},
\end{eqnarray}
In order to obtain a difference equation only with sites $n=\kappa m$, $\psi_{\kappa m-1}$ and $\psi_{\kappa m+1}$ need to
be replaced by $\psi_{\kappa (m\pm1)}$ and $\psi_{\kappa m}$ , somehow.
Rewrite Eqs. \eqref{LEGs} with transfer matrix $T$,
\begin{eqnarray}
\begin{pmatrix}
\psi_{\kappa (m-1)}\\ \psi_{\kappa(m-1)+1}\end{pmatrix}= T^{\kappa-1}
 \begin{pmatrix} \psi_{\kappa m-1}\\ \psi_{\kappa m} \end{pmatrix} \label{t1}
 \end{eqnarray}
and $m\rightarrow m+1$,
\begin{eqnarray}
\begin{pmatrix}
\psi_{\kappa (m+1)}\\ \psi_{\kappa(m+1)-1}\end{pmatrix}=T^{\kappa-1}
\begin{pmatrix} \psi_{\kappa m+1}\\ \psi_{\kappa m} \end{pmatrix}, \label{t2}
 \end{eqnarray}
where $T$ is the transfer matrix and
$$T^{\kappa-1}=\begin{pmatrix}
 E &  -1\cr
1 & 0\end{pmatrix} ^{\kappa-1} = \begin{pmatrix}
 a_{\kappa} &  -a_{\kappa-1}\cr
a_{\kappa-1} & -a_{\kappa-2} \end{pmatrix} ,$$
with
\begin{equation}
 a_{\kappa}  = \frac{1}{\sqrt{E^2-4}}((\frac{E+\sqrt{E^2-4}}{2})^{\kappa}-(\frac{E-\sqrt{E^2-4}}{2})^{\kappa}).
\end{equation}
From Eq.  \eqref{t1} and  Eq. \eqref{t2}, we get
\begin{eqnarray}
\psi_{\kappa (m-1)}  &=& a_{\kappa}  \psi_{\kappa m-1} - a_{\kappa-1}\psi_{\kappa m}; \ \label{u1}\\
\psi_{\kappa (m+1)} &=& a_{\kappa}  \psi_{\kappa m+1} - a_{\kappa-1}\psi_{\kappa m}.  \label{u2}
\end{eqnarray}
The wave functions $\psi_{\kappa m-1}$ and $\psi_{\kappa m+1}$ thus can be represented as
\begin{eqnarray}
 \psi_{\kappa m-1} &=&\frac{1}{a_{\kappa} }  \psi_{\kappa (m-1)} + \frac{a_{\kappa-1}}{a_{\kappa} } \psi_{\kappa m} \ \label{p1}\\
 \psi_{\kappa m+1} &=&\frac{1}{a_{\kappa} }  \psi_{\kappa (m+1)} + \frac{a_{\kappa-1}}{a_{\kappa} } \psi_{\kappa m}  \label{p2}
\end{eqnarray}
Substituting Eq. \eqref{p1} and Eq. \eqref{p2} into Eq. \eqref{eigen_km} yields
\begin{eqnarray}
\psi_{\kappa (m-1)} +\psi_{\kappa (m+1)} + \lambda a_{\kappa} \Delta_{m} \psi_ {\kappa m}=(E-2a_{\kappa-1})\psi_{\kappa m},\notag \\\label{eigen_km}
\end{eqnarray}
which is a decoupled difference equation at sites $n=\kappa m$.


\begin{thebibliography}{99}
\bibitem{anderson1958absence} P. W. Anderson, Absence of diffusion in
certain random lattices, Phys. Rev. \textbf{109}, 1492(1958).

\bibitem{abrahams1979scaling} E. Abrahams, P. W. Anderson, D. C.
Licciardello, and T. V. Ramakrishnan, Scaling theory of localization:
Absence of quantum diffusion in two dimensions, Phys. Rev. Lett. \textbf{42}, 673 (1979).

\bibitem{Thouless72} D J Thouless, A relation between the density of states and range
of localization for one dimensional random
systems, J. Phys. C: Solid State Phys. {\bf 5}, 77 (1972).

\bibitem{lee1985disordered} P. A. Lee and T. V. Ramakrishnan, Disordered
electronic systems, Rev. Mod. Phys. \textbf{57}, 287(1985).

\bibitem{evers2008anderson} F. Evers and A. D. Mirlin, Anderson transitions,
Rev. Mod. Phys. \textbf{80}, 1355 (2008).

\bibitem{biddle2011localization} J. Biddle, D. J. Priour, B. Wang, and S.
Das Sarma, Localization in one-dimensional lattices with
non-nearest-neighbor hopping: Generalized Anderson and Aubry- Andr\'e
models, Phys. Rev. B \textbf{83}, 075105 (2011).

\bibitem{biddle2010predicted} J. Biddle and S. Das Sarma, Predicted mobility
edges in one-dimensional incommensurate optical lattices: An exactly
solvable model of Anderson localization, Phys. Rev. Lett. \textbf{104},
070601 (2010).

\bibitem{ganeshan2015nearest} S. Ganeshan, J. H. Pixley, and S. Das Sarma,
Nearest neighbor tight binding models with an exact mobility edge in one
dimension, Phys. Rev. Lett. \textbf{114}, 146601 (2015).

\bibitem{li2016quantum} X. P. Li, J. H. Pixley, D. L. Deng, S. Ganeshan, and
S. Das Sarma, Quantum nonergodicity and fermion localization in a system
with a single-particle mobility edge, Phys. Rev. B \textbf{93}, 184204
(2016).

\bibitem{li2017mobility} X. Li, X. P. Li, and S. Das Sarma, Mobility edges
in one-dimensional bichromatic incommensurate potentials, Phys. Rev. B
\textbf{96}, 085119 (2017).

\bibitem{li2018mobility} X. Li and S. Das Sarma, Mobility edge and
interme-diate phase in one-dimensional incommensurate lattice potentials,
Phys. Rev. B \textbf{101}, 064203 (2020). 

\bibitem{DengX} X. Deng, S. Ray, S. Sinha, G. V. Shlyapnikov, and L. Santos,
One-Dimensional Quasicrystals with Power-Law Hopping, Phys. Rev. Lett.
\textbf{123}, 025301 (2019).


\bibitem{sarma1988mobility} S. Das Sarma, S. He, and X. C. Xie, Mobility
edge in a model one-dimensional potential, Phys. Rev. Lett. \textbf{61},
2144 (1988).

\bibitem{sarma1990localization} S. Das Sarma, S. He, and X. C. Xie,
Localization, mobility edges, and metal-insulator transition in a class of
one-dimensional slowly varying deterministic potentials, Phys. Rev. B
\textbf{41}, 5544 (1990).

\bibitem{YCWang2020} Y. Wang, X. Xia, L. Zhang, H. Yao, S. Chen, J. You, Q.
Zhou, and X. Liu, One dimensional quasiperiodic mosaic lattice with exact
mobility edges, Phys. Rev. Lett. \textbf{125}, 196604 (2020).


\bibitem{Ceccatto} H. A. Ceccatto, Quasiperiodic Ising Model in a Transverse
Field: Analytical Results, Phys. Rev. Lett. {\bf 62}, 203 (1989).

\bibitem{Zhou2013} L. Zhou, H. Pu, and W. Zhang, Anderson localization of
cold atomic gases with effective spin-orbit interaction in a quasiperiodic
optical lattice, Phys. Rev. A \textbf{87}, 023625 (2013).

\bibitem{Kohmoto2008} M. Kohmoto and D. Tobe, Localization problem in a
quasiperiodic system with spin-orbit interaction, Phys. Rev. B \textbf{77},
134204 (2008).

\bibitem{Cai} X. Cai, L.-J. Lang, S. Chen, and Y. Wang, Topological
superconductor to Anderson localization transition in one-Dimensional
incommensurate lattices, Phys. Rev. Lett. \textbf{110}, 176403 (2013).

\bibitem{DeGottardi} W. DeGottardi, D. Sen, and S. Vishveshwara, Majorana
fermions in superconducting 1D systems having periodic, quasiperiodic, and
disordered Potentials, Phys. Rev. Lett. \textbf{110}, 146404 (2013).

\bibitem{Chong2015} F. Liu, S. Ghosh, and Y. D. Chong, Localization and
Adiabatic Pumping in a Generalized Aubry-Andre-Harper
Model, Phys. Rev. B \textbf{91}, 014108 (2015).

\bibitem{Chandran} A. Chandran and C. R. Laumann, Localization and Symmetry Breaking in the Quantum Quasiperiodic Ising Glass,
 Phys. Rev. X \textbf{7}, 031061 (2017).

\bibitem{WangYC-review} Y.-C. Wang, X.-J. Liu and S. Chen.
Properties and applications of one dimensional quasiperiodic lattices.
Acta Physica Sinica, {\bf 68} 040301, (2019).

\bibitem{Kohmoto1983} M. Kohmoto, Metal-insulator transition and scaling for
incommensurate systems, Phys. Rev. Lett. \textbf{26}, 1198 (1983).

\bibitem{Aubry1980} S. Aubry and G. Andr\'{e}, Analyticity breaking and
Anderson localization in incommensurate lattices, Ann. Israel Phys. Soc.
\textbf{3}, 133 (1980).

\bibitem{Thouless1988} D. J. Thouless, Localization by a potential with
slowly varying period, Phys. Rev. Lett. \textbf{61}, 2141(1988).

\bibitem{roati2008} G. Roati, C. DErrico, L. Fallani, M. Fattori, C. Fort,
M. Zaccanti, G. Modugno, M. Modugno, and M. Inguscio, Anderson localization
of a non-interacting bose Ceinstein condensate, Nature (London) \textbf{453}%
, 895 (2008).

\bibitem{luschen2018} H. P. L\"uschen, S. Scherg, T. Kohlert, M. Schreiber,
P. Bordia, X. Li, S. Das Sarma, and I. Bloch, Single-particle mobility edge
in a one-dimensional quasiperiodic optical lattice, Phys. Rev. Lett. \textbf{120}, 160404 (2018).


 \bibitem{An2018} F. A. An, E. J. Meier, and B. Gadway, Engineering a flux-dependent mobility edge in
 disordered zigzag chains, Phys. Rev. X \textbf{8}, 031045 (2018).

 \bibitem{An2021} F. A. An, K. Padavi\'{c}, E. J. Meier, S. Hegde, S. Ganeshan, J. H. Pixley,
 S. Vishveshwara, and B. Gadway, Observation of tunable mobility edges in generalized Aubry-Andr\'{e}
 lattices, Phys. Rev. Lett. 126, 040603 (2021).



\bibitem{Dwiputra2022} D. Dwiputra, F. P. Zen,  Single-particle mobility edge without disorder,
Phys. Rev. B \textbf{105}, L081110 (2022).


\bibitem{Gong2021} L.-Y. Gong, H. Lu, and W.-W. Cheng, Exact Mobility Edges in 1D
Mosaic Lattices Inlaid with Slowly Varying Potentials, Adv. Theory Simul. \textbf{4}, 2100135 (2021).

\bibitem{Liu2021} Y. Liu, Y. Wang, X.-J. Liu, Q. Zhou, and S. Chen, Exact mobility edges,
PT-symmetry breaking, and skin effect in one- dimensional non-Hermitian quasicrystals,
Phys. Rev. B \textbf{103}, 014203 (2021).


\bibitem{Avila2015} A. Avila, Global theory of one-frequency Schr\"{o}inger
operators, Acta. Math. \textbf{1}, 215, (2015).

\bibitem{Avila2017} A. Avila, J. You , Q. Zhou, Sharp phase transitions for
the almost Mathieu operator, Duke. Math. J. \textbf{14}, 166 (2017).




\bibitem{Jitomirskaya1999} S. Y. Jitomirskaya, Metal-insulator transition
for the almost mathieu operator, Ann. Math. \textbf{3}, 150 (1999).












\end{thebibliography}
\end{document}